\documentclass[a4paper,12pt,dvips]{article}
\usepackage[dvips]{graphicx}

\newcounter{point}

\newcommand{\h}{^3\mathrm{H}}
\newcommand{\he}{^3\mathrm{He}}
\newcommand{\li}{^7\mathrm{Li}}

\newcommand{\pol}{\mathcal{P}}
\newcommand{\xq}{(x,Q^2)}

\newcommand{\ms}{\overline{\mathrm{MS}}}

\begin{document}

\begin{flushright}
TPJU 16/2000 \\
hep-ph/0012293
\end{flushright}

\vspace{0.5cm}

\begin{center}
\LARGE{\textbf{Nuclear effects on the spin-dependent structure
functions}} \\
\vspace{1cm}
\large{\textsc{Aleksander Sobczyk}}\footnote{~E-mail address:
asobczyk@if.uj.edu.pl} \\
\vspace{0.2cm}
\footnotesize{\emph{Institute of Physics, Jagiellonian University,
ul.~Reymonta 4, 30-059 Cracow, Poland}} \\
\vspace{0.7cm}
\large{\textsc{Jerzy Szwed}}\footnote{~E-mail address:
szwed@if.uj.edu.pl, corresponding author.}$^,$\footnote{~Work
supported by the Polish State Committee for Scientific Research
(grant No.~2 P03B 061 16).} \\
\vspace{0.2cm}
\footnotesize{\emph{Institute of Computer Science, Jagiellonian
University, ul.~Reymonta 4, 30-059 Cracow, Poland}} \\
\end{center}

\vspace{0.5cm}

\begin{abstract}
We address the question how the spin-dependent nucleon structure
function $g_1\xq$ gets modified when the nucleon is bound inside
a nucleus. We analyze the influence of nuclear interactions using
the $\Delta-\pi$ model, known to describe well the unpolarized
effect, and the free polarized parton distributions. The results
for the neutron in $^3\mathrm{He}$ and proton in $^3\mathrm{H}$,
$^7\mathrm{Li}$ and $^{19}\mathrm{F}$ are presented, showing
significant changes in the parton spin distributions and in their
moments. Scattering processes off polarized $^7\mathrm{Li}$ are
suggested which could justify these theoretical calculations and
shed more light on both nuclear spin structure and short distance QCD.
\end{abstract}

\vspace{0.3cm}

\begin{flushleft}
\end{flushleft}


\newpage

\setcounter{point}{1}
\setcounter{equation}{0}
\noindent
\textbf{\thepoint.} The influence of nuclear effects on the nucleon
structure functions received enormous interest after the measurement
by the European Muon Collaboration \cite{Aubert} of the ratio of
structure functions:
\begin{equation}
R\xq=\frac{F^{N/\mathrm{Fe}}_2\xq}{F^{N/\mathrm{D}}_2\xq},
\end{equation}
followed by a series of experiments \cite{Bodek} confirming the
nontrivial changes of the parton densities due to the nuclear
environment. Many theoretical models have described correctly
the Bjorken $x$ and $A$-dependence \cite{Arneodo} starting sometimes
from quite different assumptions. It is therefore not easy to tell
what is the underlying dynamics of the ``EMC effect''. In the search
for new tests a polarized version of the effect was proposed some
time ago \cite{Szwed_1}. In this paper we take into account
theoretical and experimental progress in the knowledge of the
nucleon spin.

We will study corrections to the spin-dependent structure function
$g^{N}_1\xq$ (where $N$ denotes proton or neutron) by calculating
deviation from unity of the ratio:
\begin{equation}
\label{r_def}
R^{A}_{\uparrow}\xq=\frac{g^{N/A}_1\xq}{g^{N}_1\xq}.
\end{equation}
Such quantity has become important due to the use of light nuclei
(D, $^3\mathrm{He}$) in the extraction of the neutron structure
function $g^{n}_1\xq$ (SMC \cite{SMC}, E143 \cite{E143}, E154
\cite{E154}, HERMES \cite{HERMES}). The definition of $g^{N/A}_1\xq$
requires some attention. It is the structure function of a single,
polarized nucleon inside the nucleus. One should keep in mind that
polarizing the nucleus in general results in a complicated spin
combinations of nucleons and it is thus not easy to extract
$g^{N/A}_1\xq$. Only in selected cases and to certain approximation
the polarization of the nucleus is equivalent to the polarization
of the nucleon in the same direction. The nuclei we have chosen
below ($^3\mathrm{He}$, $^3\mathrm{H}$ and $^7\mathrm{Li}$,
$^{19}\mathrm{F}$) are good examples of such situation.

\vspace{12pt}

\stepcounter{point}
\setcounter{equation}{0}
\noindent
\textbf{\thepoint.} The model which describes correctly the $x$
and $A$-dependence of the unpolarized EMC effect was proposed in
Ref.~\cite{Szwed_2}. Here we extend it to the polarized case.

We recall conventional picture of nucleus as a system of nucleons
and non-nucleonic objects: $\Delta$ isobars \cite{Szwed_3} and
excess pions \cite{Llewellyn} and assume that polarized deep-inelastic
scattering may occur either from constituents of the nucleons or
from constituents of the $\Delta$ isobars. Excess pions are spinless and
do not contribute directly in polarized scattering. The momentum
distributions of non-nucleonic objects come from standard nuclear
physics calculations \cite{Friman, CDAtti} and are in this sense
independent of the model. Such construction of the model has an
important advantage: we do not have to worry what the proposed
mechanism does to low energy nuclear physics.

Another important assumption is that nucleons, $\Delta$ isobars and
pions contribute incoherently to the structure function of the
nucleus. Thanks to it we can write the ``effective nucleon''
structure function in the nucleus as a sum of convolutions of
isolated hadron structure functions with momentum distributions
taken from nuclear physics
\begin{equation}
\label{general_equation}
g^{N/A}_1\xq=\int_x^A\frac{dz}{z}f^N(z)g^{N}_1\left(\frac{x}{z},Q^2\right)+\int_x^A\frac{dz}{z}f^\Delta(z)g^\Delta_1\left(\frac{x}{z},Q^2\right),
\end{equation}
where $Q^2$ is the negative momentum transfer and
\begin{equation}
z=A\frac{k^\alpha_+}{k^A_+},\quad\alpha=N,\Delta,\pi
\end{equation}
denotes light-cone $\left(k_+=k_0+k_\parallel\right)$ momentum fraction
per nucleon of the interacting nucleon, $\Delta$ isobar or pion. The
distribution functions $f^\alpha(z)$ satisfy following sum rules:
\begin{eqnarray}
\label{nucleon_sum_rule}
\int_0^Adz~f^N(z)&=&1-\langle n_\Delta\rangle, \\
\label{delta_sum_rule}
\int_0^Adz~f^\Delta(z)&=&\langle n_\Delta\rangle, \\
\label{pion_sum_rule}
\int_0^Adz~f^\pi(z)&=&\langle n_\pi\rangle,
\end{eqnarray}
as well as the momentum conservation law
\begin{equation}
\label{momentum_conservation}
\sum_\alpha\int_0^Adz~zf^\alpha(z)=\sum_\alpha\langle z_\alpha\rangle=1.
\end{equation}
Eqs.~(\ref{nucleon_sum_rule}--\ref{delta_sum_rule}) represent baryon
number conservation law, with  $\langle n_\Delta\rangle$ and
$\langle n_\pi\rangle$ defined as average numbers of $\Delta$ isobars
and excess pions in nucleus, respectively.

Let us precisely discuss each contribution entering
Eq.~(\ref{general_equation}).

\vspace{6pt}

\noindent
\textbf{\thepoint.1.} In the parton model our fundamental quantity
of interest, the spin-dependent nucleon structure function $g^N_1$, can
be expressed in terms of differences between the number densities of
quarks with spin parallel and anti-parallel to the longitudinally
polarized parent nucleon:
\begin{equation}
g^N_1(x)=\frac{1}{2}\sum_q^{N_f} e^2_q\left\{\Delta q^N(x)+\Delta\overline{q}^N(x)\right\},
\end{equation}
where
\begin{equation}
\Delta q^N(x)=q^N_\uparrow-q^N_\downarrow,\ \Delta \overline{q}^N(x)=\overline{q}^N_\uparrow-\overline{q}^N_\downarrow.
\end{equation}
In the leading-order (LO) QCD above functions become $Q^2$-dependent.
In the next-to-leading order (NLO) $g^N_1$ can be written as:
\begin{eqnarray}
g^N_1(x,Q^2) & = & \frac{1}{2}\sum_q^{N_f}e^2_q\Bigg\{\Delta q^N\xq+\Delta\overline{q}^N\xq+ \nonumber \\
&+&\left.\frac{\alpha_s\left(Q^2\right)}{2\pi}\left[\Delta C_q\otimes\left(\Delta q^N+\Delta\overline{q}^N\right)+\frac{1}{N_f}\Delta C_g\otimes\Delta g\right]\xq\right\},
\end{eqnarray}
where $\Delta g$ is polarized gluon density
($\Delta g=g_\uparrow-g_\downarrow$), $\alpha_s\left(Q^2\right)$ is
QCD running coupling constant and $N_f$ denotes the number of active
flavors. The spin-dependent Wilson coefficients $\Delta C_q$ and
$\Delta C_g$ in the $\ms$ scheme can be found in Ref.~\cite{Mertig}.
Convolution $\otimes$ is defined as usual by
\begin{equation}
\left(f\otimes g\right)(x)\equiv\int_x^1\frac{dz}{z}f(z)g\left(\frac{x}{z}\right).
\end{equation}

In $\he$ with the wave function entirely in $S$ state, two protons have
opposite spins and all the spin is carried by neutron. But in realistic
$\he$ nucleus higher partial waves ($S'$ and $D$) in the ground state wave
function lead to the spin depolarization. Effective polarization of neutron
in $\he$ is estimated in \cite{CDAtti2} to decrease to $\pol_n=(86\pm2)\%$,
whereas effective polarization of single proton is $\pol_p=(-2.8\pm0.4)\%$.
Thanks to these quantities one can write the spin-dependent structure
function $g_1^N\xq$ from Eq.~(\ref{general_equation}) as:
\begin{equation}
\label{neutron_depolarization}
g_1^N\xq=\pol_ng_1^n\xq+2\pol_pg_1^p\xq
\end{equation}
for neutron in $\he$ and as:
\begin{equation}
\label{proton_depolarization}
g_1^N\xq=\pol_ng_1^p\xq+2\pol_pg_1^n\xq
\end{equation}
for proton in $\h$. In the case of $\li$ we assumed that it consists of
$^4\mathrm{He}$, where two protons and two neutrons have opposite spins,
and of $\h$, in which we treat the spin depolarization like in
Eq.~(\ref{proton_depolarization}).

\vspace{6pt}

\noindent
\textbf{\thepoint.2.} For nuclei with $A>6$ we use an approximate formula
\cite{Llewellyn} reproducing the effect of Fermi motion:
\begin{equation}
\label{f_Li}
f^N(z)=\frac{3}{4}\left(\frac{m_N}{k_F}\right)^3\left[\left(\frac{k_F}{m_N}\right)^2-(z-1)^2\right]
\end{equation}
\begin{equation}
\label{f_domain}
\mathrm{for}\ -\frac{k_F}{m_N}\le z-1\le \frac{k_F}{m_N},
\end{equation}
and $f^N(z)=0$ otherwise. The possible corrections to this distribution
turn out to be of little importance in our problem. The Fermi momenta
$k_F$ for various nuclei can be found in Ref.~\cite{Bodek_Ritchie}.
Here $m_N$ is the nucleon mass.

The distribution $f^N(z)$ for $^3\mathrm{He}$ has been extracted from
Ref.~\cite{CDAtti} and assumed to describe also $^3\mathrm{H}$. It is
worth to note that one can calculate $f^N(z)$ from nucleon momentum
distribution $\rho_N(\vec k)$ (easily accessible in terms of
conventional nuclear theory) using approximate relationship \cite{Berger}:
\begin{equation}
f^N(z)=\int d^3\vec k~\rho_N(\vec k)\delta\left(z-\frac{k_\parallel+\sqrt{\vec k^2+m^2_N}}{m_N}\right).
\end{equation}

\vspace{6pt}

\noindent
\textbf{\thepoint.3.} Even though the pions are spinless and do not
directly enter Eq.~(\ref{general_equation}), their influence comes
trough the sum rule (\ref{momentum_conservation}) since baryons share
the momentum with pions. Effectively, this requires replacement
(in Eq.~(\ref{f_Li}) and next) of $z-1$ by $z-1+\langle z_\pi\rangle$,
where $\langle z_\pi\rangle$ is average momentum carried by pions
(all distributions are then peaked at $1-\langle z_\pi\rangle$).

\vspace{6pt}

\noindent
\textbf{\thepoint.4.} The $\Delta$ isobar structure function
$g^\Delta_1\xq$, required in Eq.~(\ref{general_equation}), is not
known from experiment. A phenomenological construction has been
presented in Ref.~\cite{Szwed_3} and can be extended to the
spin-dependent case in a straightforward way. We start from writing
the valence part of proton and neutron polarized structure functions
as:
\begin{equation}
g_{1v}^p\xq=\frac{1}{2}\left(\frac{4}{9}\Delta u_v\xq+\frac{1}{9}\Delta d_v\xq\right)=\frac{1}{2}\left(\frac{4}{9}A_0\xq+\frac{2}{9}A_1\xq\right)
\end{equation}
\begin{equation}
g_{1v}^n\xq=\frac{1}{2}\left(\frac{1}{9}\Delta u_v\xq+\frac{4}{9}\Delta d_v\xq\right)=\frac{1}{2}\left(\frac{1}{9}A_0\xq+\frac{1}{3}A_1\xq\right)
\end{equation}
where $A_I$, $I=0,1\ $ denotes the contribution in which the valence
quark is struck by the virtual photon and the remaining (spectator)
valence quarks are in spin and isospin $I$ configuration. The
$A_I\xq$ can be expressed by the valence quark distributions in the
proton:
\begin{equation}
A_0\xq=\Delta u_v\xq-\frac{1}{2}\Delta d_v\xq,
\end{equation}
\begin{equation}
A_1\xq=\frac{3}{2}\Delta d_v\xq.
\end{equation}

The $\Delta$ structure function is constructed assuming the
universality of the functions $A_I\xq$ in ground state baryons.
Noting that the spectator valence quarks in the $\Delta$ isobar
are always in spin-isospin 1 state, one writes:
\begin{equation}
g_{1v}^{\Delta^{++}}\xq=\frac{4}{9}A_1\xq=\frac{2}{3}\Delta d_v\xq,
\end{equation}
\begin{equation}
g_{1v}^{\Delta^+}\xq=\frac{1}{3}A_1\xq=\frac{1}{2}\Delta d_v\xq,
\end{equation}
\begin{equation}
g_{1v}^{\Delta^0}\xq=\frac{2}{9}A_1\xq=\frac{1}{3}\Delta d_v\xq,
\end{equation}
\begin{equation}
g_{1v}^{\Delta^-}\xq=\frac{1}{9}A_1\xq=\frac{1}{6}\Delta d_v\xq.
\end{equation}
In addition we assume that the sea quarks and gluons remain in the
same shape in any of the $\Delta$ isobars and neglect Fermi motion
effects for the $\Delta$ isobars in our analysis. Hence $f^\Delta(z)$
has form:
\begin{equation}
f^\Delta(z)=\langle n_\Delta\rangle\delta\left(z-1+\langle z_\pi\rangle\right).
\end{equation}

\vspace{12pt}

\stepcounter{point}
\setcounter{equation}{0}
\noindent
\textbf{\thepoint.} To demonstrate nuclear effects on the
spin-dependent structure functions $g^p_1$ and $g^n_1$, we have
chosen three recent parametrizations of free polarized parton
distributions in the nucleon at the next-to-leading order (NLO)
in the $\overline{\mathrm{MS}}$ scheme: AAC \cite{AAC}
(called AAC-NLO-2), LSS \cite{LSS} and TBK \cite{TBK}. For
completeness, we also give predictions of our model using
leading-order (LO) parametrization \cite{AAC} (called AAC-LO).
Nuclear parameters required in the calculation are extracted from
Ref.~\cite{Friman}. For $^3\mathrm{He}$ and $^3\mathrm{H}$ they
take the values: $\langle n_\Delta\rangle=0.02$,
$\langle n_\pi\rangle=0.05$, $\langle z_\pi\rangle=0.038$,
whereas for $^7\mathrm{Li}$: $\langle n_\Delta\rangle=0.04$,
$\langle n_\pi\rangle=0.09$, $\langle z_\pi\rangle=0.069$. The Fermi
momentum $k_F$ for $^7\mathrm{Li}$ is $0.86\ \mathrm{fm}^{-1}$.
In Eq.~(\ref{general_equation}) $g^\Delta_1$ stands for averaged
over isospin spin-dependent $\Delta$ isobar structure function.

The results are presented at $Q^2_0=1\ \mathrm{GeV}^2$. The first
moment of the spin-dependent structure function $g^\beta_1\xq$
(where $\beta=p,\ p/A\ ,n\ ,n/A$) is defined as:
\begin{equation}
\Gamma^\beta_1\left(Q^2\right)=\int^1_0dx~g^\beta_1\xq,
\end{equation}
and its values for free proton, free neutron and for proton and
neutron in various nuclei are presented in all considered
parametrizations in Table~\ref{full_results}.

\begin{table}[!ht]
\begin{center}
\caption{First moments $\Gamma^\beta_1\left(Q^2_0=1\ \mathrm{GeV}^2\right)$}
\label{full_results}
\begin{tabular}{ccccc}
\\\hline\hline
& { } AAC (NLO) { } & { } AAC (LO) { } & { } LSS (NLO) { } & { } TBK (NLO) { }\\\hline
$\Gamma^p_1$ & 0.129 & 0.144 & 0.129 & 0.113 \\
$\Gamma^{p/^3\mathrm{H}}_1$ & 0.110 & 0.123 & 0.109 & 0.095 \\
$\Gamma^{p/^7\mathrm{Li}}_1$ & 0.101 & 0.113 & 0.103 & 0.090 \\\hline
$\Gamma^n_1$ & -0.054 & -0.067 & -0.050 & -0.062 \\
{ }$\Gamma^{n/^3\mathrm{He}}_1${ } & -0.057 & -0.069 & -0.053 & -0.064 \\
\hline\hline
\end{tabular}
\end{center}
\end{table}

We expect our model to work for $0.1\le x\le 1$, since at smaller
$x$ possible shadowing effects, not included in our calculation,
may contribute significantly \cite{Frankfurt}. The question how
the first moment of $g_1$ is modified can be answered only partially
because of this limit of applicability of the model. Defining:
\begin{equation}
\Gamma^\beta_{1,\;y}\left(Q^2\right)=\int^1_ydx~g^\beta_1\xq,
\end{equation}
in Table~\ref{partial_results} we present quantities analogical to those
of 
Table~\ref{full_results}, but integrated in the region $0.1\le x\le 1$.

\begin{table}[!ht]
\begin{center}
\caption{First moments $\Gamma^\beta_{1,\;0.1}\left(Q^2_0=1\ \mathrm{GeV}^2\right)$}
\label{partial_results}
\begin{tabular}{ccccc}
\\\hline\hline
& { } AAC (NLO) { } & { } AAC (LO) { } & { } LSS (NLO) { } & { } TBK (NLO) { } \\\hline
$\Gamma^p_{1,\;0.1}$ & 0.081 & 0.097 & 0.099 & 0.091 \\
$\Gamma^{p/^3\mathrm{H}}_{1,\;0.1}$ & 0.067 & 0.080 & 0.082 & 0.075 \\
$\Gamma^{p/^7\mathrm{Li}}_{1,\;0.1}$ & 0.063 & 0.076 & 0.078 & 0.071 \\\hline
$\Gamma^n_{1,\;0.1}$ & -0.024 & -0.022 & -0.021 & -0.020 \\
{ }$\Gamma^{n/^3\mathrm{He}}_{1,\;0.1}${ } & -0.026 & -0.025 & -0.024 & -0.023 \\
\hline\hline
\end{tabular}
\end{center}
\end{table}

\vspace{12pt}

\begin{figure}[!hbp]
\begin{center}
\includegraphics[scale=0.70]{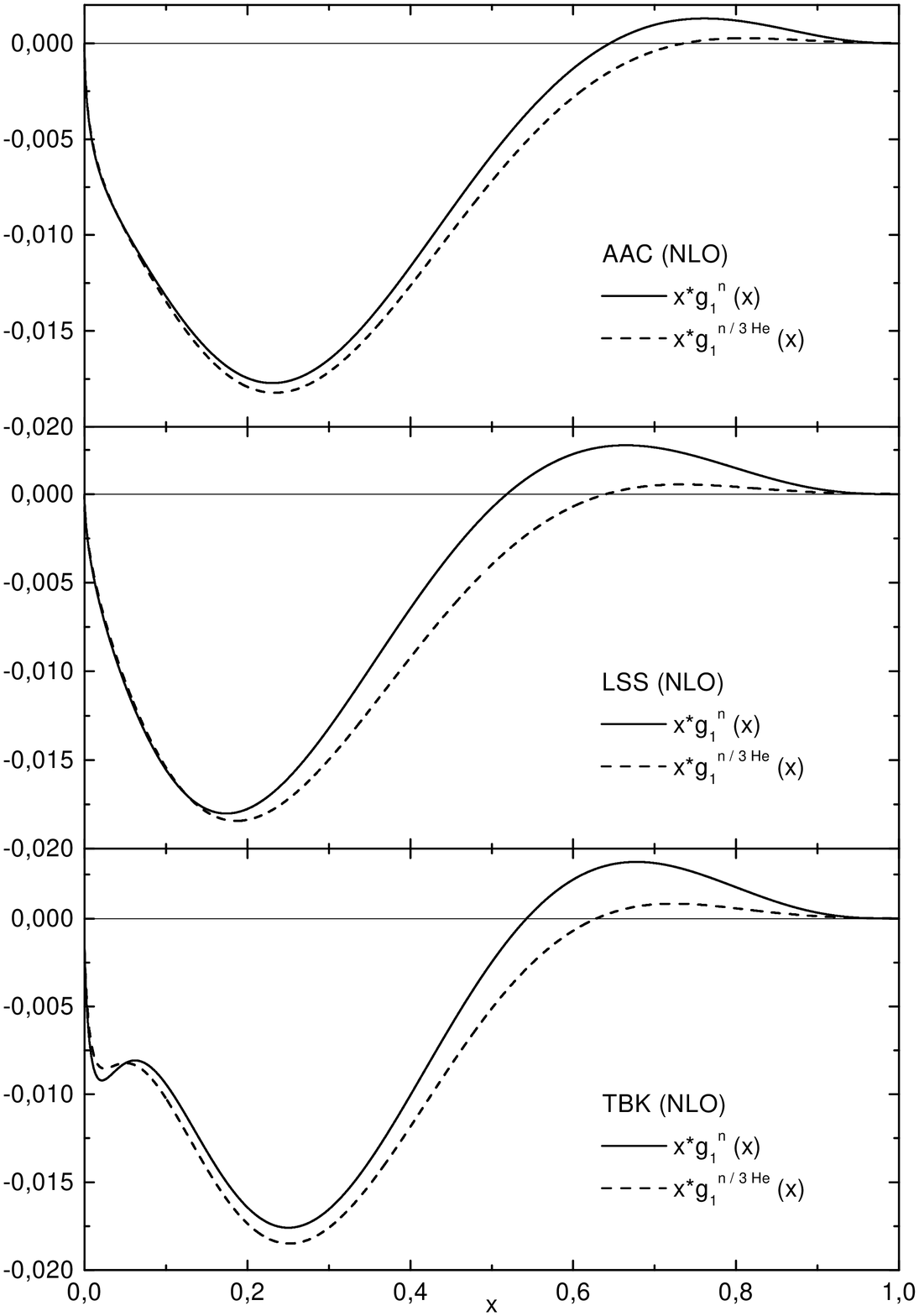}
\caption{The NLO neutron structure functions $xg_1^n(x)$ and $xg_1^{n/\he}(x)$
at $Q_0^2=1\ \mathrm{GeV}^2$ in various parametrizations}
\label{neutrons}
\end{center}
\end{figure}

\stepcounter{point}
\setcounter{equation}{0}
\noindent
\textbf{\thepoint.} We start the discussion from $^3\mathrm{He}$
nucleus, since it is an ideal target to extract the neutron structure
function $g^n_1\xq$. We plotted the spin-dependent structure functions
of neutron in $\he$ and free neutron in three different NLO parametrizations
in Fig.~\ref{neutrons}. From Table \ref{full_results} and
\ref{partial_results} one reads:
\begin{equation}
\frac{\Gamma^n_1}{\Gamma^{n/^3\mathrm{He}}_1}\approx\left\{\begin{array}{c}0.96\\0.89\end{array}\right.\quad\mathrm{for}\quad\begin{array}{c}0\ge x\ge1.0\\0.1\ge x\ge1.0\end{array},
\end{equation}
what means 11\% decrease of the first moment of the spin-dependent
structure function of neutron in $\he$ due to nuclear effects in the
range where our models gives predictions.

\begin{figure}[!hbp]
\begin{center}
\includegraphics[scale=0.70]{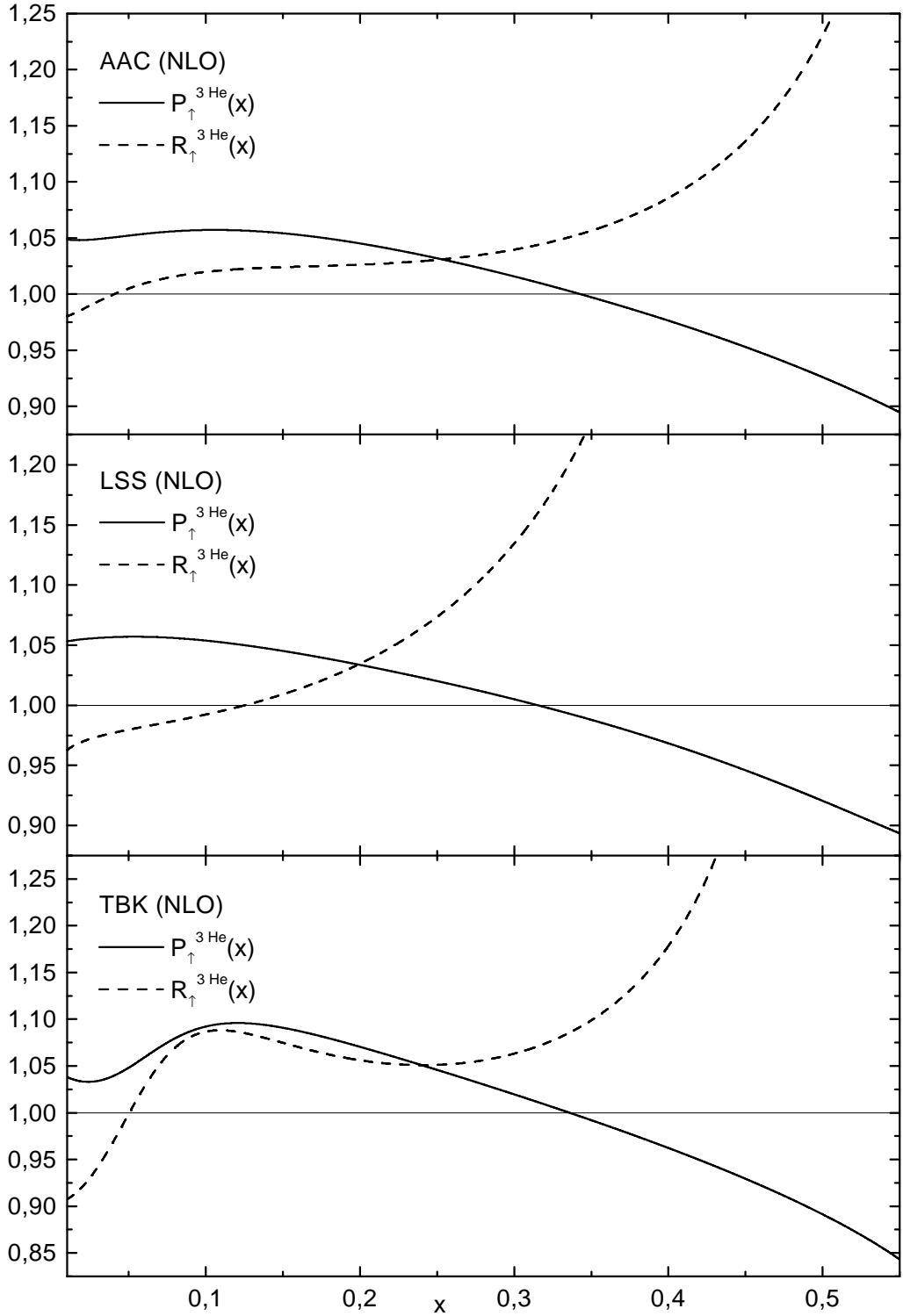}
\caption{The ratios $P_\uparrow^{\he}(x)$ and $R_\uparrow^{\he}(x)$
at $Q_0^2=1\ \mathrm{GeV}^2$ in various NLO parametrizations}
\label{nratios}
\end{center}
\end{figure}

One should remember that in the experimental analyses (like
\cite{E154,HERMES}) a correction for the spin depolarization described
by Eq.~(\ref{neutron_depolarization}) is often included. That is why it
is interesting to see what are the corrections predicted by our model
not only to the free neutron structure function $g_1^n\xq$, but also to
the function $\pol_ng_1^n\xq+2\pol_pg_1^p\xq$. Therefore in
Fig.~\ref{nratios} we presented two ratios:
\begin{equation}
P^{\he}_\uparrow\xq=\frac{g_1^{n/\he}\xq}{\pol_ng_1^n\xq+2\pol_pg_1^p\xq}
\end{equation}
and
\begin{equation}
R^{\he}_\uparrow\xq=\frac{g_1^{n/\he}\xq}{g_1^n\xq}
\end{equation}
in various NLO parametrizations. The ratio $P^{\he}_\uparrow\xq$ measures
the influence only of the $\Delta$ isobars, excess pions and Fermi motion
on the spin-dependent structure function of neutron in $\he$. If these
effects were not important, $P^{\he}_\uparrow\xq$ would be equal to unity.
The ratio $R^{\he}_\uparrow\xq$ describes all nuclear effects, including the spin depolarization in $\he$. We do not present $P^{\he}_\uparrow\xq$ and
$R^{\he}_\uparrow\xq$ in the whole $x$ region because the neutron
structure function crosses zero.

\begin{figure}[!hbp]
\begin{center}
\includegraphics[scale=0.45]{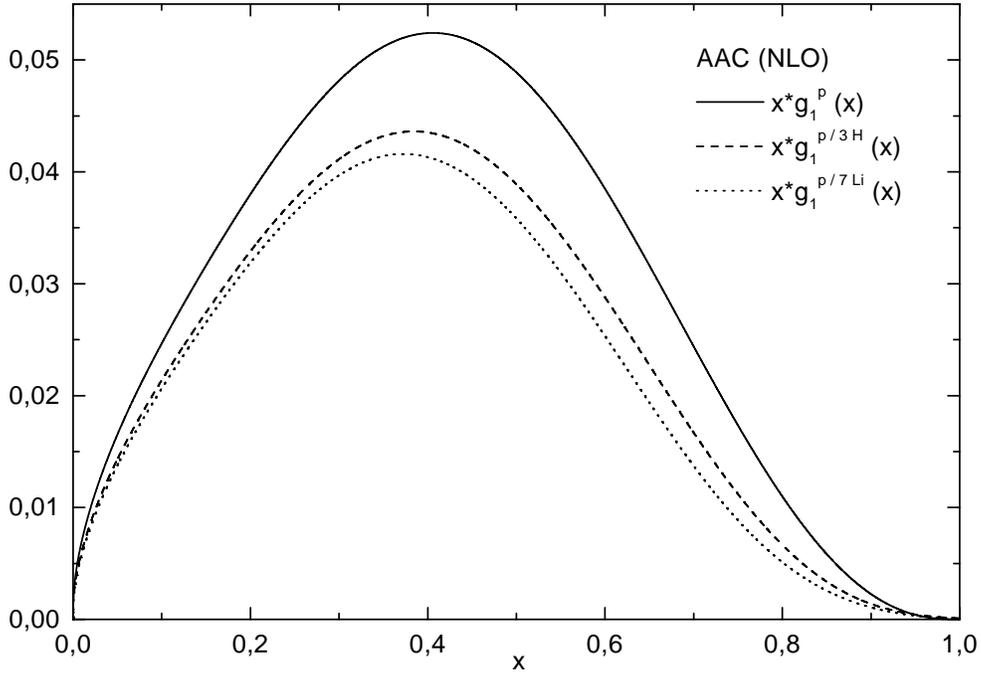}
\caption{The NLO proton structure functions $xg_1^p(x)$, $xg_1^{p/\h}(x)$
and $xg_1^{p/\li}(x)$ at $Q_0^2=1\ \mathrm{GeV}^2$ in AAC parametrization}
\label{protons}
\end{center}
\end{figure}

\begin{figure}[!hbp]
\begin{center}
\includegraphics[scale=0.45]{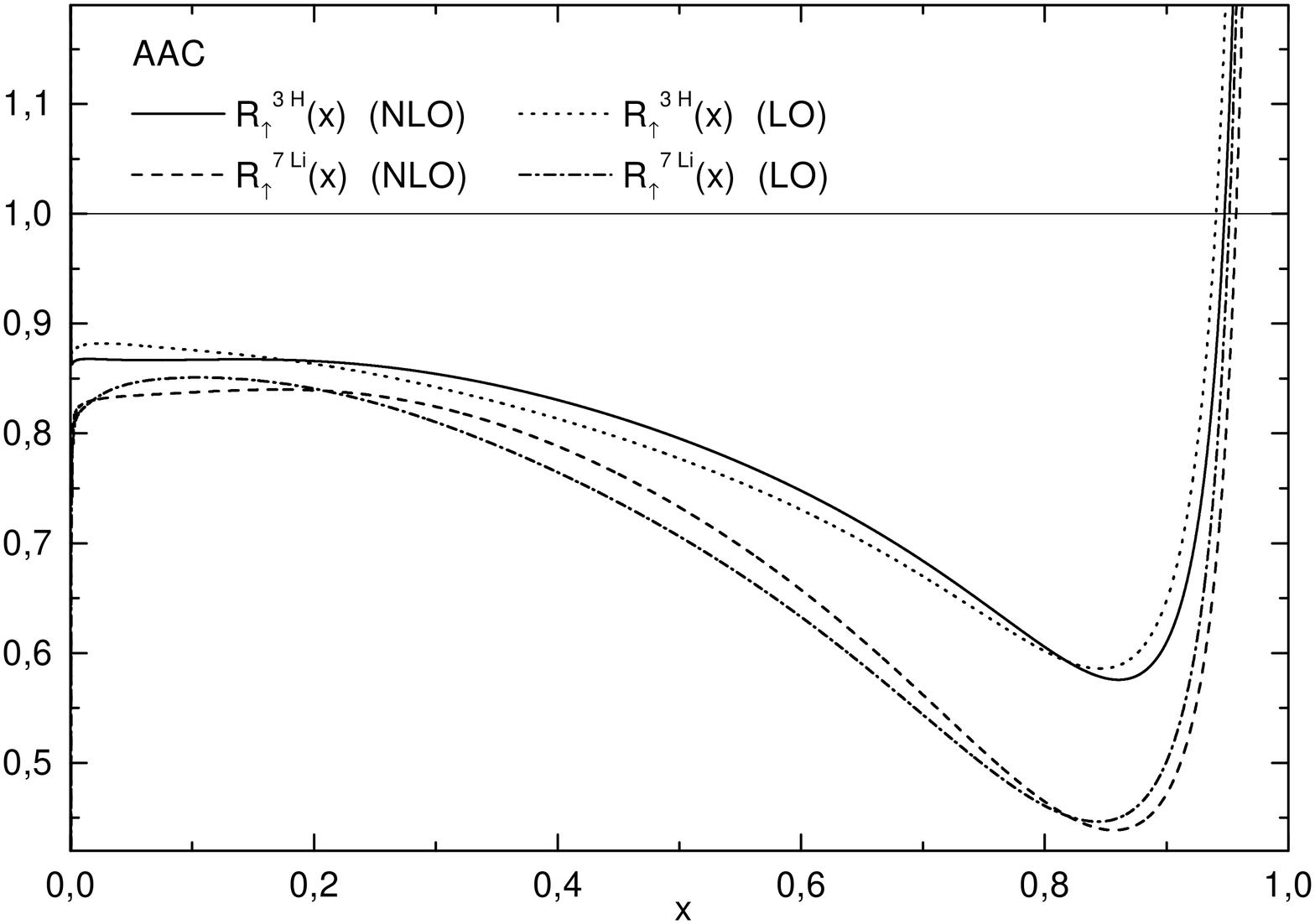}
\caption{The ratios $R^{\h}_{\uparrow}(x)$ and $R^{\li}_\uparrow$
calculated using NLO and LO AAC parametrization at
$Q_0^2=1\ \mathrm{GeV}^2$}
\label{pratios}
\end{center}
\end{figure}

The corrections to the proton structure function $g^p_1\xq$ are
studied in $^3\mathrm{H}$ and $^7\mathrm{Li}$ (we also mention
results for $^{19}\mathrm{F}$). The $^7\mathrm{Li}$ nucleus is our
best example not only for the effect which is very pronounced, but
also because this nucleus seems to be a realistic polarized target.
The $^3\mathrm{H}$ is calculated to check the Bjorken sum
rule \cite{Bjorken} for system with $A=3$. The spin-dependent structure
functions for free proton, proton in $^3\mathrm{H}$ and proton
in $^7\mathrm{Li}$ are shown in Fig.~\ref{protons} in NLO AAC
parametrization. We do not plot them in the remaining
parametrizations, since there is a very small
parametrization-dependence in our predictions for $g^{p/A}_1$.
In Fig.~\ref{pratios} ratios:
\begin{equation}
R^{\h}_\uparrow\xq=\frac{g_1^{p/\h}\xq}{g_1^p\xq}
\end{equation}
and
\begin{equation}
R^{\li}_\uparrow\xq=\frac{g_1^{p/\li}\xq}{g_1^p\xq}
\end{equation}
are shown in both NLO and LO AAC
parametrization. As compared to $^3\mathrm{He}$ richer content
of non-nucleonic objects makes the nuclear effect deeper
for $^7\mathrm{Li}$. The resulting corrections to the first
moments presented in Table 1 and 2 are also considerably larger:
\begin{equation}
\frac{\Gamma^{p/^7\mathrm{Li}}_1}{\Gamma^p_1}\approx\left\{\begin{array}{c}0.79\\0.78\end{array}\right.\quad\mathrm{for}\quad\begin{array}{c}0\ge x\ge 1\\0.1\ge x\ge1\end{array}.
\end{equation}

Having calculated the nuclear effects for both proton in
$^3\mathrm{H}$ and neutron in $^3\mathrm{He}$ we are able to check
the Bjorken sum rule for system with $A=3$. The numbers presented in
Table \ref{full_results} and \ref{partial_results} show that:
\begin{equation}
\frac{\Gamma_1^{p/\h}-\Gamma_1^{n/\he}}{\Gamma_1^p-\Gamma_1^n}\approx\left\{\begin{array}{c}0.91\\0.88\end{array}\right.\quad\mathrm{for}\quad\begin{array}{c}0\ge x\ge1\\0.1\ge x\ge1\end{array},
\end{equation}
what means 12\% reduction in the region where our model is applicable.

The results for proton in $^{19}\mathrm{F}$ are very similar to
those of $^7\mathrm{Li}$ due to the saturation of nuclear
parameters $\langle n_\Delta\rangle$, $\langle n_\pi\rangle$
and $\langle z_\pi\rangle$. This nucleus, less realistic as a
polarized target, is interesting because of possible application
in the hunt for neutralino as a dark matter candidate \cite{Ellis}.

One can certainly improve the model presented above. 
It would be interesting  to include shadowing effects
and extend the model to low $x$ region ($x<0.1$). Another 
improvement would be the inclusion of 
interference terms resulting from $N-\Delta$ transitions. 
Although it is possible to construct 
the $N-\Delta$  spin dependent structure function in analogy to the 
$\Delta$, we are unable to extract from  the nuclear matter calculation,
we are basing on \cite{Friman}, the density  $f^{N-\Delta}(z)$ in
the nucleus. Approach which attributes all nuclear effects (except
depolarization Eqs.~(\ref{neutron_depolarization}--\ref{proton_depolarization})
to the interference term \cite{Boros} is conceptually different from
ours: the absolute normalisation of this term is there a free parameter.

To summarize, we recall the idea how the nucleon spin-dependent
structure functions get modified due to nuclear environment. The
model we have used to present the effect has not been chosen by
accident. Among other advantages it can be extended from
unpolarized to polarized version without new, fundamental
assumptions. Whereas the case of $^3\mathrm{He}$ serves rather only
as a warning what size of corrections should one expect when extracting
the neutron structure function from polarized $^3\mathrm{He}$ target
experiments (our predictions for neutron in $\he$ are within
experimental error bars), the $^7\mathrm{Li}$ nucleus seems to be more
promising. With present experimental techniques one may seriously
think of deep inelastic polarized lepton - polarized
$^7\mathrm{Li}$ scattering or polarized hadron - polarized
$^7\mathrm{Li}$ scattering with direct photon or muon pair
production. In all cases the modification due to nuclear effects
should be measurable. The expected results are interesting for
both nuclear structure and QCD studies. One should keep in mind
that what is usually measured in deep inelastic scattering is
the asymmetry:
\begin{equation}
A_1\xq\simeq\frac{g_1\xq}{F_1\xq},
\end{equation}
where $F_1$ is the unpolarized structure function. Since the
nuclear effect is similar on both the numerator and denominator,
one may be misled by the fact that the asymmetry itself shows
no significant change as compared to the free nucleon case.

\vspace{12pt}

\noindent
\textbf{Acknowledgements} \\
The authors would like to thank Bogus\l{}aw Kamys, John Millener
and Larry Trueman for discussions. 



\newpage

\newpage

\begin{thebibliography}{99}
\bibitem{Aubert} EMC Collaboration, J.J.Aubert \textit{et al.}, Phys.~Lett.~\textbf{B123}, 275 (1983);
\bibitem{Bodek} A.Bodek \textit{et al.}, Phys.~Rev.~Lett.~\textbf{50}, 1431 (1983); \textbf{51}, 534 (1983); \\ R.G.Arnold \textit{et al.}, Phys.~Rev.~Lett.~\textbf{52}, 727 (1984); \\ NMC Collaboration, P.Amaudruz \textit{et al.}, Z.~Phys.~\textbf{C51}, 387 (1991); \textbf{C53}, 73 (1992);
\bibitem{Arneodo} for review see: M.Arneodo, Phys.~Rept.~\textbf{240}, 301 (1994); \\ K.J.Heller and J.Szwed, Acta Phys.~Pol.~\textbf{B16}, 157 (1985);
\bibitem{Szwed_1} L.deBarbaro, K.J.Heller and J.Szwed, Jagiellonian University preprint TPJU 24/84 (1984); \\ L.deBarbaro, K.J.Heller and J.Szwed, J.~Phys.~Soc.~Jpn~\textbf{55} Suppl., 962 (1986) (Proc.~VI Int.~Symp.~Polar.~Phenom.~in Nucl.~Phys., Osaka 1985);
\bibitem{SMC} SMC Collaboration, B.~Adeva \emph{et al.}, Phys.~Rev.~\textbf{D58}, 112002 (1998);
\bibitem{E143} E143 Collaboration, K.~Abe \emph{et al.}, Phys.~Rev.~\textbf{D58}, 112003 (1998);
\bibitem{E154} E154 Collaboration, K.~Abe \emph{et al.}, Phys.~Rev.~Lett.~\textbf{79}, 26 (1997);
\bibitem{HERMES} HERMES Collaboration, K.~Ackerstaff \emph{et al.}, Phys.~Lett.~\textbf{B404}, 383 (1997);
\bibitem{Szwed_2} J.Kubar, G.Plaut and J.Szwed, Z.~Phys.~\textbf{C23}, 195 (1984);
\bibitem{Szwed_3} J.Szwed, Phys.~Lett.~\textbf{B128}, 245 (1983);
\bibitem{Llewellyn} C.H.Llewellyn Smith, Phys.~Lett.~\textbf{B128}, 107 (1983); \\ M.Ericson and A.W.Thomas, Phys.~Lett.~\textbf{B128}, 112 (1983);
\bibitem{Friman} B.L.Friman, V.R.Pandharipande and R.B.Wiringa, Phys.~Rev.~Lett.~\textbf{51}, 763 (1983);
\bibitem{CDAtti} C.Ciofi degli Atti and S.Simula, Phys.~Rev.~\textbf{C53}, 1689 (1996);
\bibitem{Mertig} R.Mertig and W.L.~van Neerven, Z.~Phys.~\textbf{C70}, 637 (1996); \\ W.Vogelsang, Phys.~Rev.~\textbf{D54}, 2023 (1996);
\bibitem{CDAtti2} C.~Ciofi degli Atti \emph{et al.}, Phys.~Rev.~\textbf{C48}, R968 (1993);
\bibitem{Bodek_Ritchie} A.Bodek and J.L.Ritchie, Phys.~Rev.~\textbf{D23}, 1070 (1981);
\bibitem{Berger} E.L.Berger, F.Coester and R.B.Wiringa, Phys.~Rev.~\textbf{D29}, 398 (1984);
\bibitem{AAC} AAC Collaboration, Y.Goto \textit{et al.}, Phys.~Rev.~\textbf{D62}, 034017 (2000);
\bibitem{LSS} E.Leader, A.V.Sidorov and D.B.Stamenov, Phys.~Lett.~\textbf{B462}, 189 (1999);
\bibitem{TBK} S.Tatur, J.Bartelski and M.Kurzela, Acta Phys.~Pol.~\textbf{B31}, 647 (2000);
\bibitem{Frankfurt} L.Frankfurt, V.Guzey and M.Strikman, Phys.~Lett.~\textbf{B381}, 379 (1996);
\bibitem{Bjorken} J.D.Bjorken, Phys.~Rev.~\textbf{148}, 1467 (1966);
\bibitem{Ellis} J.Ellis and R.A.Flores, Nucl.~Phys.~\textbf{B400}, 25
(1993);
\bibitem{Boros} C.Boros \emph{et al.}, preprint hep-ph/0008064, ADP-00-30/T413 (2000).
\end{thebibliography}
\end{document}